\documentclass[aps,floatfix,showpacs]{revtex4}
\usepackage{amsmath}
\usepackage{amsfonts}
\usepackage{amssymb}
\usepackage{graphics,epsfig}
\usepackage{graphicx}
\begin{document}

\title{Generating static spherically symmetric anisotropic solutions of Einstein's equations from isotropic Newtonian solutions}
\author{Kayll Lake \cite{email}}
\affiliation{Department of Physics, Queen's University, Kingston,
Ontario, Canada, K7L 3N6 }
\date{\today}

\begin{abstract}
I use the Newtonian equation of hydrostatic equilibrium for an isotropic fluid sphere to generate exact anisotropic solutions of Einstein's equations. The input function is simply the density. An infinite number of regular solutions are constructed, some of which satisfy all the standard energy conditions. Two classes of these solutions generalize the Newtonian polytropes of index $0$ and $1$.
\end{abstract}
\pacs{04.20.Cv, 04.20.Jb, 04.40.Dg}
\maketitle
\section{Introduction}
The study of exact spherically symmetric perfect fluid solutions of Einstein's equations has been reinvigorated over the last decade or so and we now have a remarkable collection of solution generating techniques \cite{visser}. At the same time, interest in exact anisotropic fluid solutions of Einstein's equations has grown \cite{anisotropic}. Whereas a solution generating technique is known for anisotropic fluids \cite{herrera}, this requires the specification of two input functions and one is not guaranteed a reasonable output, say in the density $\rho$. A less general technique is given here, but the algorithm presented requires only one input function, the density itself. The technique is demonstrated by the construction of an infinite number of regular solutions, some of which satisfy all the standard energy conditions.
\section{Geometry}

Let us write the geometry in the familiar form \cite{notation}
\begin{equation}\label{metric}
    ds^2=\frac{dr^2}{1-2m(r)/r}+r^2d\Omega^2-e^{2\Phi(r)}dt^2
\end{equation}
where $d\Omega^2$ is the metric of a unit 2-sphere ($d\theta^2+sin^2\theta d\phi^2$). The source of (\ref{metric}), by way of Einstein's equations, is taken to be a comoving fluid described by the stress-energy tensor $T^{\alpha}_{\beta}=diag[p(r),P(r),P(r),-\rho(r)]$. Einstein's equations give the effective gravitational mass
\begin{equation}\label{mass}
    m=4 \pi \int _{0}^{r}\rho(x) x^2dx,
\end{equation}
the source equation for $\Phi$
\begin{equation}\label{phi}
    \Phi^{'}=\frac{m+4 \pi r^3 p}{r(r-2m)},
\end{equation}
and the generalized Tolman-Oppenheimer-Volkoff equation which we write in the form
\begin{equation}\label{peqn}
    P=\frac{r}{2}(p^{'}+(\rho+p) \Phi^{'})+p,
\end{equation}
where $'\equiv d/dr$. The fluid distribution is assumed to terminate at $r=R$ where $p(R)=0$ and to join there, by way of a regular boundary surface, onto a Schwarzschild vacuum of mass $m(R)\equiv M$. No further conditions need be specified for this junction.

\bigskip

A fluid is considered regular if it is free of singularities. Scalar polynomial singularities involve scalar invariants built out of the Riemann tensor. Due to the spherical symmetry assumed here, there are only four independent invariants of this type \cite{invar}. A direct calculation shows that $\rho^{'}$ and $P^{'}$ do not enter these invariants and that the Ricci invariants are regular as long as $\rho, p$ and $P$ are. However, the second Weyl invariant grows like
\begin{equation}\label{weyl}
   \left(\frac{rp^{'}+2(p-P)}{\rho-3p+4P}\right)\frac{1}{r^6}
\end{equation}
as $r \rightarrow 0$. As a consequence, we have the following necessary conditions for regularity
\begin{equation}\label{regular}
    p^{'}(0)=0, \;\;\;\;p(0)=P(0),
\end{equation}
the latter of which is already obvious from (\ref{peqn}). One could, of course, incorporate derivatives of the Riemann tensor into the construction of invariants. For example, a calculation shows that $\square \mathcal{R}$ grows like
\begin{equation}\label{boxr}
    \frac{\rho^{'}-2 P^{'}}{r}
\end{equation}
as $r \rightarrow 0$ where $\mathcal{R}$ is the Ricci scalar and $\square$ is the covariant d'Alembertian. However, the background theory involves the Riemann tensor, not its derivatives, and so we impose no a priori restrictions beyond (\ref{regular}).
\section{The Algorithm}
The algorithm proposed here involves the use of the Newtonian hydrostatic equilibrium equation for an isotropic fluid,
\begin{equation}\label{hydro}
    p^{'}=-\frac{m \rho}{r^2}.
\end{equation}
In effect then $\rho$ and $p$ are related exactly as in the Newtonian theory of isotropic fluids, e.g. \cite{chandra}. A solution in Einstein's theory is obtained simply by reading off $P$ from (\ref{peqn}). No isotropic solutions for Einstein's theory can be found in this way since if we set $P=p$ in (\ref{peqn}), and use (\ref{hydro}), the resultant quadratic equation for $p$ returns $p(0)\leq 0$.  One can take the view that it is the tangential stress $P$ that holds these configurations together. From (\ref{peqn}) and (\ref{hydro}), and assuming $\rho \geq 0$, we have
\begin{equation}\label{inequality}
    P \geq p \geq 0.
\end{equation}
Explicitly, inserting (\ref{phi}) and (\ref{hydro}) into (\ref{peqn}) we obtain
\begin{equation}\label{explicit}
    P=-\frac{1}{2}\frac{m \rho}{r}+\frac{1}{2}\frac{(\rho+p)(m+4 \pi r^3p)}{r-2m}+p.
\end{equation}
 As a result, the weak and classical strong energy conditions are automatically satisfied via the proposed algorithm, as long as $\rho \geq 0$. The dominant energy condition is not \cite{energy}. However, in the present circumstance, the classical strong energy conditions are in fact weaker than the weak energy conditions and so the trace energy condition
\begin{equation}\label{trace}
    2P+p\leq \rho
\end{equation}
is usually applied. Subject to (\ref{trace}) it is known that the Buchdahl bound $R/2M \geq 9/8$ \cite{buchdahl} is robust \cite{andre}. From (\ref{trace}) we can write
\begin{equation}\label{radtio}
    \frac{P}{\rho} \leq \epsilon
\end{equation}
and consider the dominant energy condition as $\epsilon=1$ and the stronger trace energy condition as $\epsilon=1/2$ \cite{ivanov}.
\section{Examples}
The procedure outlined above can be executed with the specification of $\rho$ alone. We demonstrate this here with an infinite number of simple examples mostly derived from the ansatz
\begin{equation}\label{rhos}
    \rho=-ar^n+b
\end{equation}
where $a$ and $b$ are constants $>0$ and $n$ is an integer $\geq 1$. First, let us consider the simplest case, that of constant density, $a=0$ \cite{constant}.
\subsection{$a=0$}
\begin{figure}[ht]
\epsfig{file=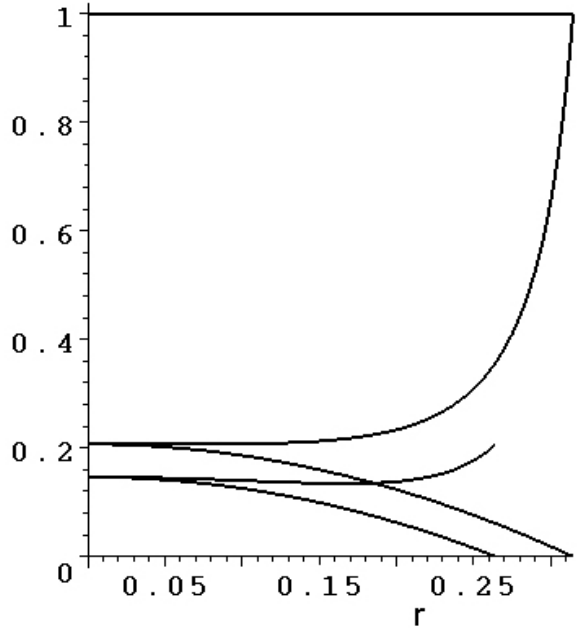,height=3.5in,width=3in,angle=0}
\caption{\label{figure1} The variables $P$ (increasing) and $p$ (decreasing to $0$ at $r=R$) for the critical case $R=R_{c}$ and for $R=R_{c}-0.05$ (for which $R/2M \sim 1.707$) for the constant density solutions with $b=1$.}
\end{figure}

\begin{figure}[ht]
\epsfig{file=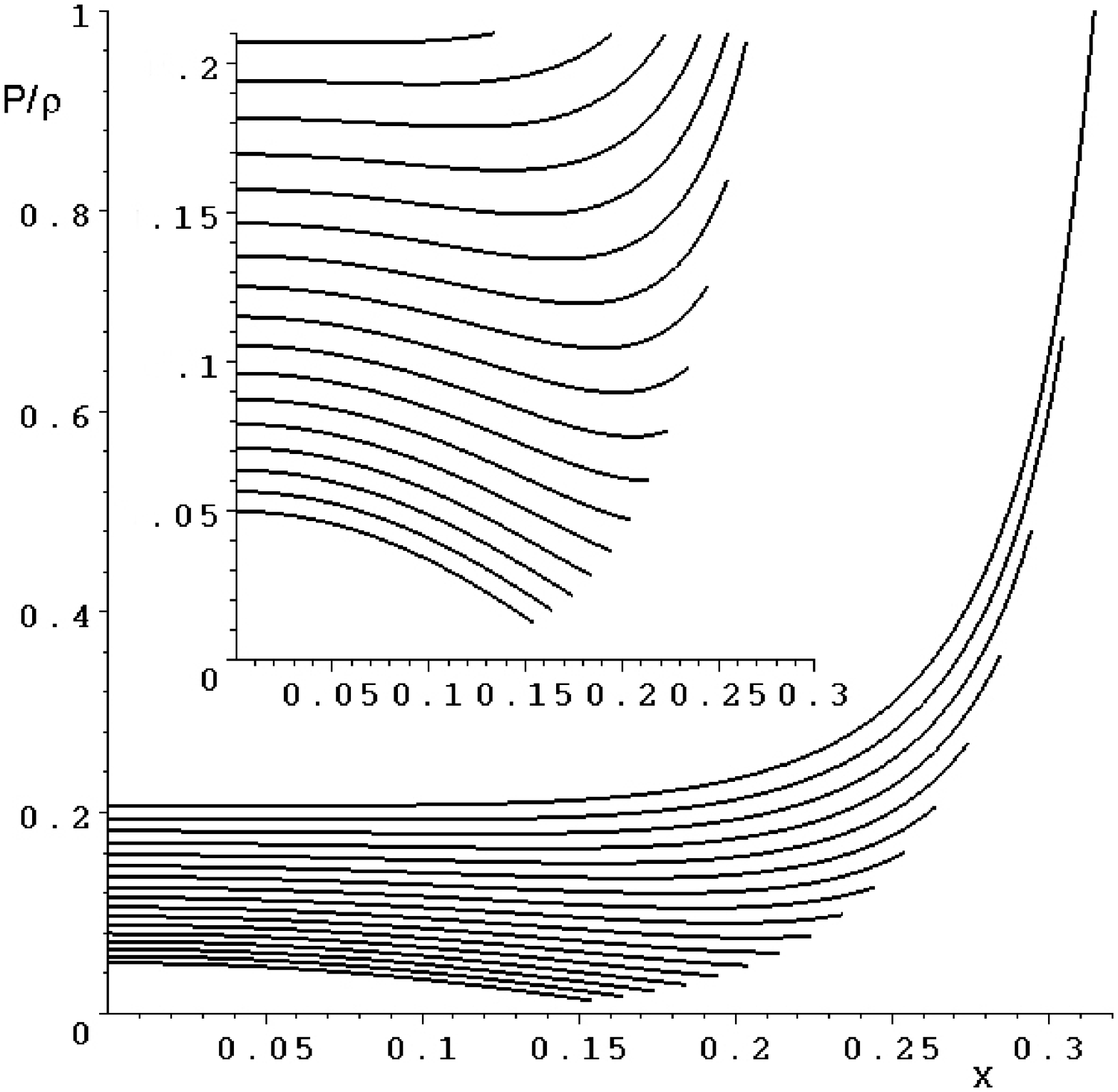,height=3.5in,width=3in,angle=0}
\caption{\label{figure2} The ratio $P/\rho$ for models characterized by $X=X_{c}-\delta$ where $\delta=0$ for the top curve and $\delta$ increases by $0.01$ for each subsequent curve. The insert shows more detail. For these models $R/2M=3/8 \pi (X_{c}-\delta)^2$.}
\end{figure}

 From (\ref{mass}) and (\ref{rhos}) it follows that $m=4 \pi b r^3/3$ and so $M=4 \pi b R^3/3$. From (\ref{hydro}), and setting $p(R)=0$, we obtain
 \begin{equation}\label{pconst}
    p=\frac{2}{3}\pi b^2\left( R-r \right)  \left( R+r \right)
 \end{equation}
 and so $p^{'}=-4 \pi b^2 r/3$.
This is a Newtonian polytrope of zero index, e.g. \cite{chandra}. We can now simply read off $P$ from (\ref{peqn}) with (\ref{phi}). We find
\begin{widetext}
\begin{equation}\label{Psconst}
    P=\frac{2 \pi b^2}{3(3-8 \pi b r^2)}\left({r}^{2} ( 4\,{\pi }^{2}{b}^{2}{r}
^{4}-3+8\,\pi \,b{r}^{2} ) -{R}^{2} ( 8\,{\pi }^{2}{b}^{2}{
r}^{4}-4\,{\pi }^{2}{b}^{2}{r}^{2}{R}^{2}-3 )\right).
\end{equation}
\end{widetext}
To avoid singularities in $P$ then
\begin{equation}\label{singconst}
    R< \sqrt{\frac{3}{8 \pi b}}
\end{equation}
and so for non-singular solutions $R<2M$ as expected.
At $r=0$ we find
\begin{equation}\label{centralconditionsconst}
    \rho(0)=b,\;\rho^{'}(0)=\;p^{'}(0)=P{'}(0)=0
\end{equation}
along with
\begin{equation}\label{centralpPconst}
    p(0)=P(0)=\frac{2 \pi b^2 R^2}{3}.
\end{equation}
At $r=R$ we find
\begin{equation}\label{boundaryconditionsconst}
    \rho(R)=b,\;\rho^{'}(R)=p(R)=0,\;p^{'}(R)=-\frac{4 \pi b^2}{3}R
\end{equation}
along with
\begin{equation}\label{Pgradboundaryconst}
    P(R)=\frac{16 \pi^2 b^3}{3}\left(\frac{R^4}{3-8 \pi b R^2}\right).
\end{equation}
For solutions with $P(R)\leq \rho$, since $P(r) \geq 0$, the dominant energy condition is also satisfied throughout the solution. From (\ref{Pgradboundaryconst}) for these solutions we find
\begin{equation}\label{limit}
    R\leq R_{c} \equiv \left(\left(\frac{3}{4 \pi b}\right)(\sqrt{2}-1)\right)^{1/2},
\end{equation}
where $P(R_{c})=\rho$, and so in the limiting case $R_{c}/2M=1/2(\sqrt{2}-1)\sim 1.207$ somewhat larger that the the Buchdahl bound $R/2M=9/8=1.125$ \cite{buchdahl}. Some examples are shown in Figure \ref{figure1}.
In order to demonstrate that energy conditions of the type (\ref{radtio}) can be satisfied, it is convenient to define $x^2 \equiv b r^2$ and $X^2=b R^2$ so that
\begin{equation}\label{Psconstmod}
    \frac{P}{\rho}=\frac{2 \pi }{3(3-8 \pi {x}^2)}\left({x}^{2} ( 4\,{\pi }^{2}{x}
^{4}-3+8\,\pi \,{x}^2 ) -{X}^{2} ( 8\,{\pi }^{2}{
x}^{4}-4\,{\pi }^{2}{x}^{2}{X}^{2}-3 )\right).
\end{equation}
A number of examples are shown in Figure \ref{figure2}.
\bigskip

\subsection{$a>0$}
We set
\begin{equation}\label{Rs}
    R=\left(\frac{b}{a}\right)^{1/n}
\end{equation}
so that $\rho(R)=0$. Other boundary conditions are given below. From (\ref{mass}) and (\ref{rhos}) it follows that
\begin{equation}\label{masss}
    m=\frac{4 \pi}{3}r^3 \left(\frac{(3+n)b-3ar^n}{3+n}\right)
\end{equation}
and so
\begin{equation}\label{Ms}
    M=\frac{4 \pi}{3}R^3 \left(\frac{nb}{3+n}\right).
\end{equation}
From (\ref{hydro}), (\ref{rhos}) and (\ref{masss}) we obtain
\begin{widetext}
\begin{equation}\label{ps}
    p=\frac{2 \pi}{3}\left({\frac { \,  -{r}^{2} \left( {b}^{2} \left( 3+n \right)
 \left( 2+n \right)  \left( 1+n \right) +a{r}^{n} \left( 3\,a{r}^{n}
 \left( 2+n \right) -2\,b \left( 6+n \right)  \left( 1+n \right)
 \right)  \right) +{R}^{2}{b}^{2}{n}^{2} \left( 4+n \right)   }
{ \left( 3+n \right)  \left( 2+n \right)  \left( 1+n \right) }}\right).
\end{equation}
\end{widetext}
We can now simply read off $P$ from (\ref{peqn}) with (\ref{phi}). We find
\begin{equation}\label{Ps}
    P=\left(\frac{2 \pi}{3 (3+n)(2+n)^2(1+n)^2}\right)\left(\frac{H(r)}{\mathcal{P}(r)}\right)
\end{equation}
where $H$ is a polynomial of degree $6+4n$, given explicitly in Appendix A, and
\begin{equation}\label{Poly}
    \mathcal{P}=3(3+n)-8 \pi (3+n)br^2+24 \pi a r^{2+n}.
\end{equation}
As shown in Appendix B, (\ref{Poly}) has no roots for $r>0$ as long as
\begin{equation}\label{alimit}
    a>\left(\frac{3+n}{8 \pi}\right)\left(\frac{16 \pi b}{3(2+n)}\right)^{n/2+1}\left(\frac{n}{2}\right)^{n/2} .
\end{equation}
Subject to condition (\ref{alimit}) then the solutions given here are free of singularities throughout the range $0 \leq r \leq R$. Whereas the configurations satisfy the weak and strong energy conditions, the dominant energy condition is satisfied for sufficiently large $a$. This is demonstrated below.

\bigskip

At $r=0$ we find
\begin{equation}\label{centralconditions}
    \rho(0)=b,\;\rho^{'}(0)=-anr^{n-1},\;p^{'}(0)=P{'}(0)=0
\end{equation}
along with
\begin{equation}\label{centralpP}
    p(0)=P(0)=\left(\frac{2 \pi}{3}\right)\left({\frac { \,{R}^{2}{b}^{2}{n}^{2} ( 4+n ) }{ (
3+n )  ( 2+n )  ( 1+n ) }}\right).
\end{equation}
At $r=R$ we find
\begin{equation}\label{boundaryconditions}
    \rho(R)=p(R)=P(R)=p^{'}(R)=0,\;\rho^{'}(R)=-\frac{nb}{R}
\end{equation}
along with
\begin{equation}\label{Pgradboundary}
    P^{'}(R)=\frac{16}{3}\left({\frac {{R}^{3}{n}^{3}{b}^{3}{\pi }^{2}}{(-3(3+n)+8\,\pi
\,b{R}^{2}n )  ( 3+n ) }}\right).
\end{equation}

\bigskip

From (\ref{Rs}), (\ref{Ms}) and (\ref{alimit}) we find
\begin{equation}\label{tenuity}
   \frac{R}{2M} =\frac{3(3+n)}{8 \pi n b R^2} \equiv \alpha > \left(\frac{3}{2}\right) \left( \frac {2(3+n)}{3(2+n)} \right) ^{2/n+1} >1.
\end{equation}
Since $\displaystyle\lim_{n\to\infty} \alpha = 1$, the solutions given here allow an arbitrarily large degree of compactness, in principle. However, smaller values of $\alpha$ correspond to smaller values of $a$ which lead to a violation of the dominant (and therefor trace) energy condition. Again, this is demonstrated below. From (\ref{tenuity}) it follows that the Buchdahl bound $R/2M=9/8$ is obtained for
\begin{equation}\label{buchdahl}
    a=\left(\frac{3 \pi n b^{2/n+1}}{3+n}\right)^{n/2}.
\end{equation}

\subsubsection{$n=1$}
\begin{figure}[ht]
\epsfig{file=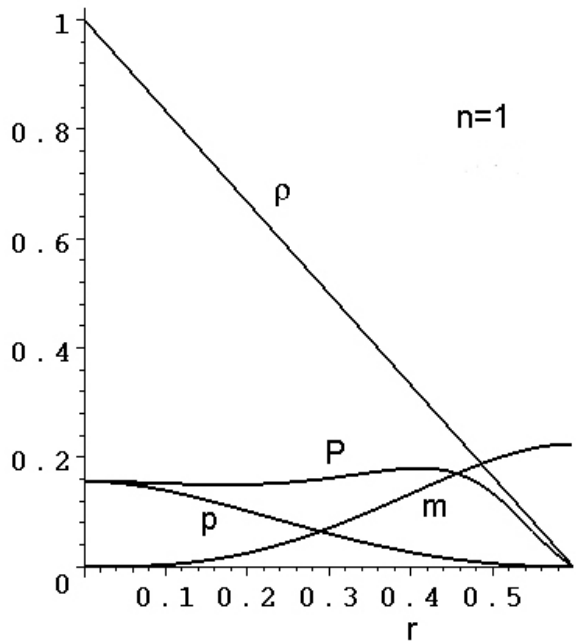,height=3.5in,width=3in,angle=0}
\caption{\label{figure3} The variables $\rho, P, p$ and $m$ for $a=2/3\sqrt{2 \pi}b^{3/2}$ and $b=1$ for the linear density solution.}
\end{figure}

\begin{figure}[ht]
\epsfig{file=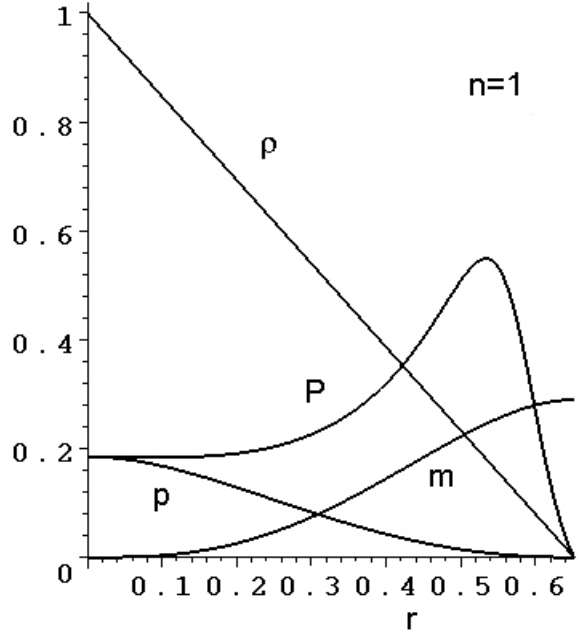,height=3.5in,width=3in,angle=0}
\caption{\label{figure4} The variables $\rho, P, p$ and $m$ in the Buchdahl limit $a=\sqrt{3/8}\sqrt{2 \pi}b^{3/2}$ and $b=1$ for the linear density solution.}
\end{figure}

For $n=1$ there is no ``core'' in the density. It follows from (\ref{alimit}) that
\begin{equation}\label{n1a}
    a > \frac{16}{27}\sqrt{2 \pi}b^{3/2},
\end{equation}
and from (\ref{tenuity}) that $R/2M > 256/243$. The physical variables are shown in Figure \ref{figure3} for $a=2/3\sqrt{2 \pi}b^{3/2}$ and in Figure \ref{figure4} for $a=\sqrt{3/8}\sqrt{2 \pi}b^{3/2}$, the Buchdahl limit. In the first case all energy conditions (excepting the trace) are satisfied and $\alpha =4/3$. (The trace energy condition can be satisfied with a slightly larger $a$, for example, $a=19/27\sqrt{2 \pi}b^{3/2}$.) In the second case the dominant energy condition fails. Unlike the case $a=0$, the minimum $\alpha$ for which the dominant energy condition is satisfied must be found numerically for each $n$. For $n=1$ the minimum $\alpha \sim 1.29$, of course above the Buchdahl limit.
\subsubsection{$n=2$}
\begin{figure}[ht]
\epsfig{file=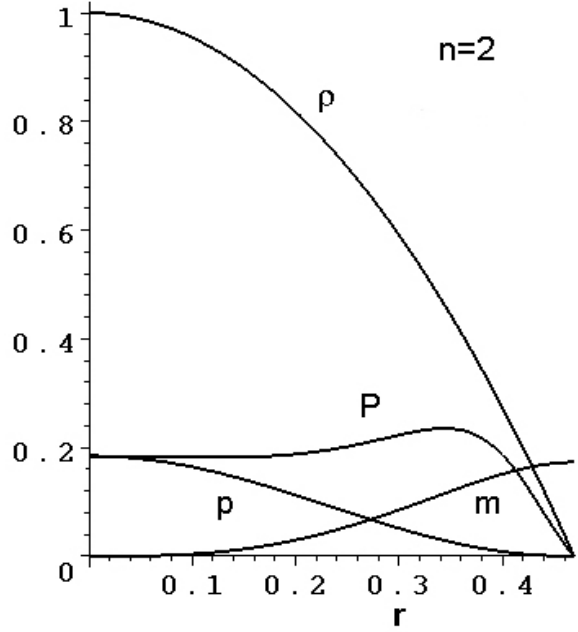,height=3.5in,width=3in,angle=0}
\caption{\label{figure5} The variables $\rho, P, p$ and $m$ for $a=13/9\pi b^{2}$ and $b=1$ for the quadratic density solution.}
\end{figure}

\begin{figure}[ht]
\epsfig{file=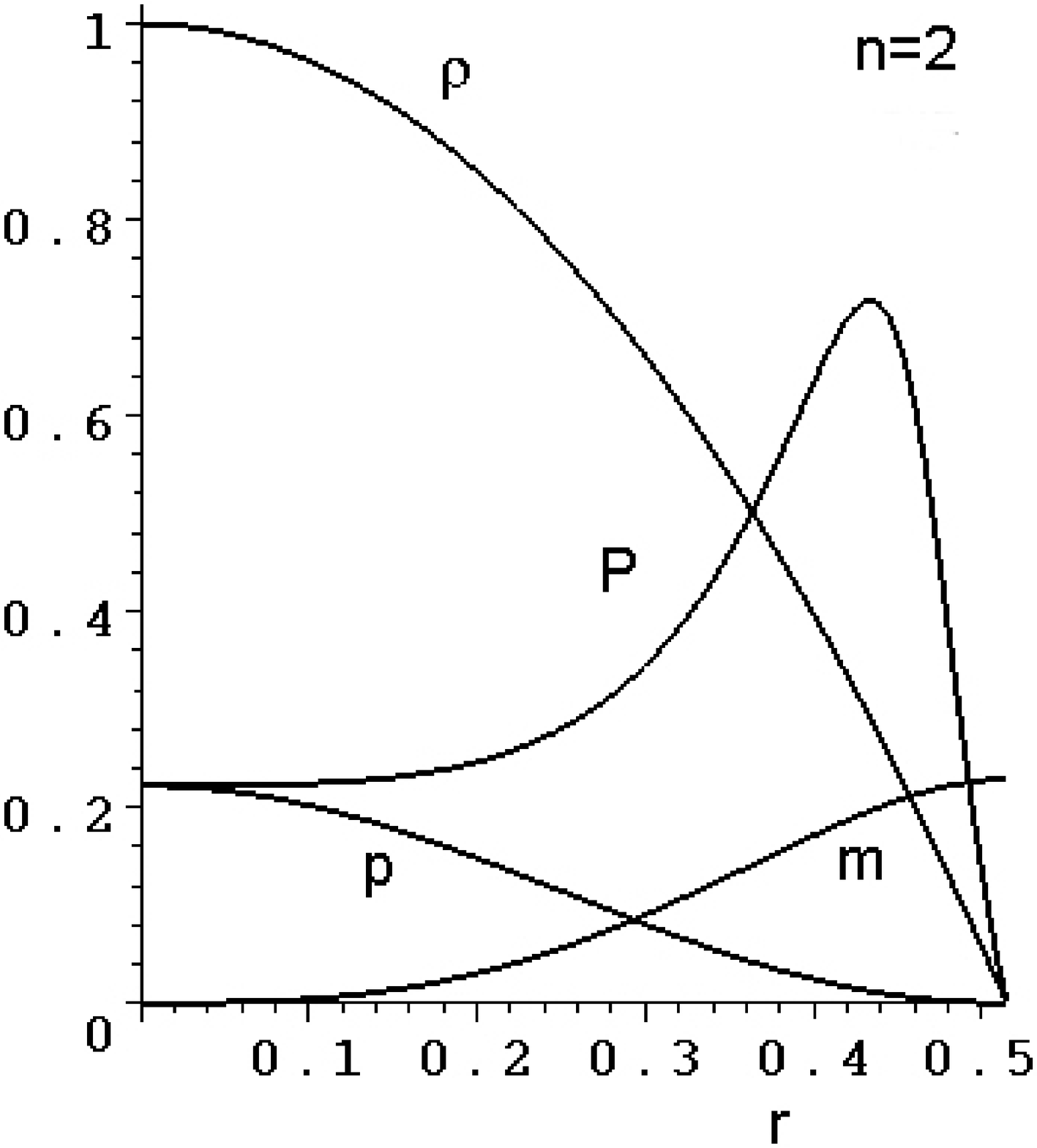,height=3.5in,width=3in,angle=0}
\caption{\label{figure6} The variables $\rho, P, p$ and $m$ in the Buchdahl limit $a=6/5\pi b^{2}$ and $b=1$ for the quadratic density solution.}
\end{figure}

\begin{figure}[ht]
\epsfig{file=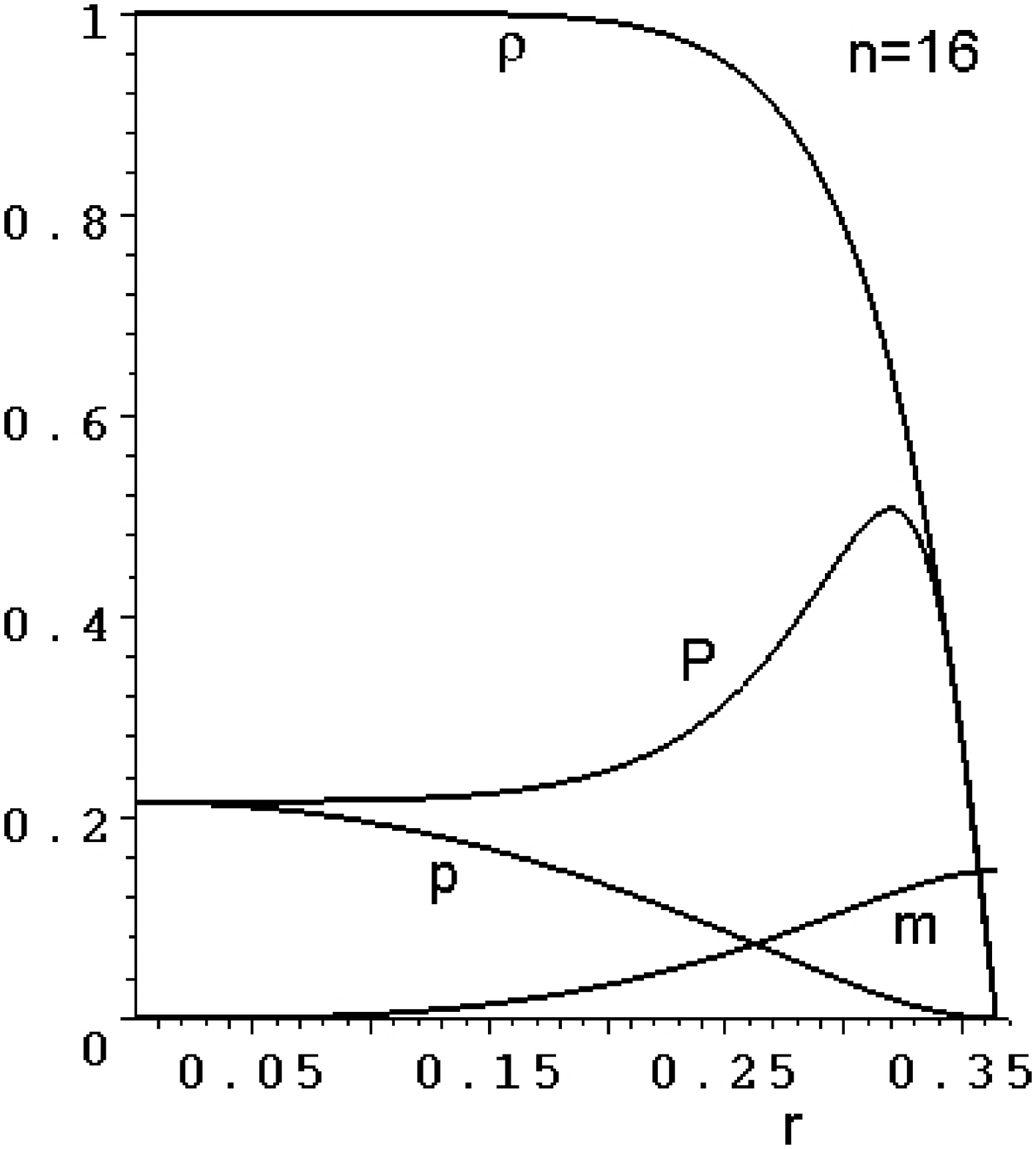,height=3.5in,width=3in,angle=0}
\caption{\label{figure7} The variables $\rho, P, p$ and $m$ for $a \sim 33 \pi^4 b^{5}$ and $b=1$ for the solution with $n=16$. This case gives the minimum $\alpha \sim 1.24$ for which the dominant energy condition is satisfied.}
\end{figure}

For $n=2$ it follows from (\ref{alimit}) that
\begin{equation}\label{n2a}
    a > \frac{10}{9} \pi b^{2},
\end{equation}
and from (\ref{tenuity}) that $R/2M > 25/24$. The physical variables are shown in Figure \ref{figure5} for $a=13/9\pi b^{2}$ and in Figure \ref{figure6} for $a=6/5\pi b^{2}$, the Buchdahl limit. In the first case all energy conditions (excepting the trace) are satisfied and $\alpha =65/48 \sim 1.354$.  (The trace energy condition can be satisfied with a slightly larger $a$, for example, $a=14/9\pi b^{2}$.) In the second case the dominant energy condition fails.
The solutions with $n>2$ are qualitatively similar, but with a flatter density profile for smaller $r$, that is, a larger ``core''. An example is shown in Figure \ref{figure7}.
\subsubsection{Polytrope of index 1}
\begin{figure}[ht]
\epsfig{file=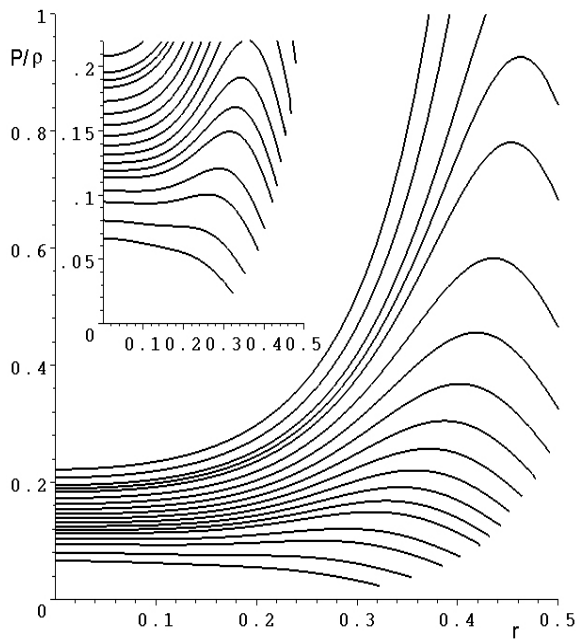,height=3.5in,width=3in,angle=0}
\caption{\label{figure8} The ratio $P/\rho$ for models characterized by $a=(3+\delta)\sqrt{\pi \rho_{0}}$ where the top curve has $\delta=0$ (the Buchdahl bound). Going down the curves, $\delta$ increases by $0.1$ except for the fourth curve down for which $\delta=0.25$ and the bottom 3 curves for which $\delta=1.6, 2$ and $2.5$ respectively. For smaller values of $\delta$ the curves are truncated. The insert shows more detail.}
\end{figure}

The Newtonian polytrope of zero index was discussed above, but it is well known that there are two finite analytical polytropic solutions, e.g. \cite{chandra}. In this section we use a classical polytrope of index 1 to generate an anisotropic fluid. Let us take
\begin{equation}\label{poly1rho}
    \rho=\rho_{0}\frac{\sin(a r)}{a r}
\end{equation}
where $a$ is a constant, so that from (\ref{hydro})
\begin{equation}\label{{poly1p}}
    p=\frac{2 \pi}{a^2}\rho^2
\end{equation}
and
\begin{equation}\label{{poly1b}}
    R=\frac{\pi}{a}.
\end{equation}
From (\ref{mass}) and  (\ref{poly1rho}) we have
\begin{equation}\label{poly1mass}
    m=\frac{4 \pi \rho_{0}}{a^3}(\sin(a r)-a r \cos(a r))
\end{equation}
and so
\begin{equation}\label{poly1M}
    M=\frac{4 \pi^2}{a^3}\rho_{0}=\frac{4}{a}\rho_{0}R^2
\end{equation}
from which we have
\begin{equation}\label{poly1a}
    \frac{R}{2M}=\frac{a^2}{8 \pi \rho_{0}}
\end{equation}
and so $a^2>9 \pi \rho_{_{0}}$ for the Buchdahl bound to hold. We now find
\begin{equation}\label{poly1pp}
    P=\frac{I(r)}{32\,\pi \,{r}^{5}{a}^{8} \left( r-2\,m \right) }
\end{equation}
where $I$ is a polynomial of degree $6$, given explicitly in Appendix C. At $r=0$ we find
\begin{equation}\label{poly1centralconditions}
    \rho(0)=\rho_{0},p(0)=P(0)=\frac{2 \pi}{a^2} \rho_{0}^2,\;\rho^{'}(0)=\;p^{'}(0)=P^{'}(0)=0.
\end{equation}
At $r=R$ we find
\begin{equation}\label{polyboundaryconditions}
    \rho(R)=p(R)=P(R)=p^{'}(R)=0,\;\rho^{'}(R)=-\frac{\rho_{0}a}{\pi}
\end{equation}
along with
\begin{equation}\label{polyPgradboundary}
    P^{'}(R)=-\frac{16 \pi \rho_{0}^3}{a(a^2-8 \pi \rho_{0})}<0
\end{equation}
where the last inequality assumes the Buchdahl bound. These models are qualitatively similar to the case $n=2$ discussed above. In a manner analogous to Figure \ref{figure2} some details associated with these models are shown in Figure \ref{figure8}.
\section{Discussion}
An algorithm which converts isotropic Newtonian static fluid spheres into anisotropic Einsteinian static fluid spheres has been given which requires one input function, the density. An infinite number of explicit regular solutions have been generated, some of which satisfy all the standard energy conditions.
\begin{acknowledgments}
It is a pleasure to thank H{\aa}kan Andr\'{e}asson and Boyko Ivanov for useful comments. This work was supported by a grant from the Natural Sciences and
Engineering Research Council of Canada. Portions of this work were made possible by use of \emph{GRTensorII} \cite{grt}.
\end{acknowledgments}
\appendix
\section{$H$}
For $a>0$ and $n \geq 1$ we find
\begin{widetext}
\begin{eqnarray}
H(r)={b}^{2}{n}^{2}{R}^{2} (n+4)[ 4\,{R}^{2}{b}^{2}{r}^
{2}{\pi }^{2}{n}^{2}( n+4 ) -8\,{b}^{2}{r}^{4}{\pi }^{2}
( n+3)( n+2)( n+1) +16\,b{r}^
{n+4}{\pi }^{2}a( n+6)( n+1)\\ \nonumber  -6\,n\pi \,{r}
^{n+2}a( n+2)( n+1) -24\,{\pi }^{2}{r}^{4+2
\,n}{a}^{2}( n+2) +3\,( n+3)( n+2
 ) ( n+1)] \\ \nonumber+{b}^{2}{r}^{2}( n+3
 ) ^{2} ( n+2) ^{2}( n+1) ^{2}( 4
\,{b}^{2}{r}^{4}{\pi }^{2}-3+8\,b{r}^{2}\pi) +6\,a b{r}^{n+2}
 ( n+3)( n+2 )( n+6)( n+
1 ) ^{2}\\ \nonumber-2\,\pi \,{b}^{2}a{r}^{n+4}( n+36)(
n+3 )( n+2) ^{2} ( n+1) ^{2}-16\,{\pi
}^{2}{b}^{3}a{r}^{n+6}( n+3 )( n+2)(
n+6 )( n+1) ^{2}\\ \nonumber-9\,{a}^{2}{r}^{2+2\,n}( n+3
)( n+1)( n+2) ^{2}+12\,{a}^{2}\pi
\,b{r}^{4+2\,n}( n+2)( 3\,{n}^{2}+20\,n+36
) ( n+1 ) ^{2}\\ \nonumber+8\,{\pi }^{2}{b}^{2}{a}^{2}{r}^{6+2
\,n} ( n+1 )( 5\,{n}^{3}+47\,{n}^{2}+144\,n+108
) -18\,{a}^{3}\pi \,{r}^{4+3\,n}( 4+3\,n )(
n+1) ( n+2 ) ^{2}\\ \nonumber-48\,{\pi }^{2}{a}^{3}b{r}^{6+3\,n
} ( n+6 ) ( n+2 )( n+1) +36\,{
\pi }^{2}{a}^{4}{r}^{6+4\,n}( n+2 ) ^{2}.
\end{eqnarray}
\end{widetext}
\section{Proof of (\ref{alimit})}
This follows immediately from the following elementary theorem \cite{levin}: For
\begin{equation}\label{levin}
    1-ax^m+bx^n
\end{equation}
$a, b$ real and $n>m>0$ integers, there are no positive roots as long as $a>0$ and
\begin{equation}\label{levin1}
    b>(\frac{a}{\alpha})^{\alpha}(\alpha-1)^{\alpha-1}
\end{equation}
where $\alpha \equiv n/m$.
\section{$I$}
In terms of (\ref{poly1rho}) and (\ref{poly1mass}) we find
\begin{widetext}
\begin{eqnarray}
I(r)=256\,{\pi }^{4}{r}^{4} \left( \rho_{{0}}-\rho \right) ^{2} \left( \rho
_{{0}}+\rho \right) ^{2}\\ \nonumber
128\, {\pi }^{3} \rho  \,{r}^{3}{a}^{2} \left( {a}^{2}{r}
^{3} \left( 3\,{\rho}^{2}+{\rho_{{0}}}^{2} \right) -4\,r \left( \rho_{
{0}}-\rho \right)  \left( \rho_{{0}}+\rho \right) +2\,  m
   \left( \rho_{{0}}-\rho \right)  \left( \rho_{{0}}+\rho
 \right)  \right)\\ \nonumber
 -32\,{\pi }^{2} {r}^{2} {a}^{4}\left( 6\,{a}^{2}  m  {\rho}^{2}{r
}^{3}+2\,{\rho}^{2}{r}^{2}-2\,{\rho_{{0}}}^{2}{r}^{2}-5\,{\rho_{{0}}}^
{2} m  r+13\,  m  {\rho}^{2}r-3\,{m}^{2}{
\rho}^{2}+{\rho_{{0}}}^{2}{m}^{2} \right)\\ \nonumber
+8\, \pi \rho  \,  m  r {a}^{6}\, \left( 3\,
{a}^{2} m  {r}^{3}+4\,{r}^{2}+14\,  m  r-2
\,{m}^{2} \right) \\ \nonumber
-{a}^{8}{m}^{2} \left( 4\,{r}^{2}+10\, m  r-{m}^{2}
 \right).
\end{eqnarray}
\end{widetext}


\begin{thebibliography}{}
\bibitem{notation} We use geometrical units and usually designate functional dependence only on the first appearance of a function. We assume that all functions are at least $C^2$.
\bibitem{constant} It is important to note that the solution we are considering is not the most general solution to (\ref{peqn}) with constant $\rho$. Rather, it is the solution, with constant $\rho$, that satisfies (\ref{hydro}). Compare, for example, R. L. Bowers and E. P. T. Liang, Astrophys. J. \textbf{188}, 657 (1974).
\end{thebibliography}
\end{document}